\documentclass[aps,twocolumn]{revtex4}
\usepackage{graphicx}
\usepackage{dcolumn}
\usepackage{bm}
\usepackage{CJK}
\usepackage{amsfonts}
\usepackage{psfrag}
\usepackage{wrapfig}
\usepackage{subfigure}
\usepackage{makeidx}
\usepackage{multirow}
\usepackage{epsf}
\usepackage{hyperref}
\usepackage{amsmath}
\usepackage{cases}

\begin{document}

\title{Different types of nonlinear localized and periodic waves in an erbium-doped fiber system}
\author{Yang Ren$^1$}
\author{Zhan-Ying Yang$^1$}\email{zyyang@nwu.edu.cn (Zhan-Ying Yang);
nwudavid@163.com (Chong Liu)}
\author{Chong Liu$^{1}$}
\author{Wen-Li Yang$^2$}
\address{$^1$School of Physics, Northwest University, Xi'an
710069, China}
\address{$^2$Institute of Modern Physics, Northwest University, Xi'an
710069, China}
\date{August 10, 2015}
\begin{abstract}
We study nonlinear waves on a plane-wave background in an erbium-doped fiber system, which is governed by the coupled nonlinear Schr\"{o}dinger and the Maxwell-Bloch equations. We find that prolific different types of nonlinear localized and periodic waves do exist in the system, including multi-peak soliton, periodic wave, antidark soliton, and W-shaped soliton (as well as the known bright soliton, breather, and rogue wave). In particular, the dynamics of these waves can be extracted from a unified exact solution, and the corresponding existence conditions are presented explicitly. Our results demonstrate the structural diversity of the nonlinear waves in this system.

\end{abstract}

\maketitle

\section{INTRODUCTION}
Nonlinear waves propagating in an erbium-doped fiber has attracted special attention, since the resonant absorption of the erbium-doped two levels system is a good solution to the fiber-optic signal attenuation problem \cite{1}. In general, the dynamic of nonlinear waves in the erbium-doped fibers is described by the coupled nonlinear Schr\"{o}dinger and the Maxwell-Bloch (NLS-MB) models \cite{1,s1,s2}. Like the standard NLS model, the standard bright (i.e., zero background) solitons \cite{s1,s2}, and the localized waves on a plane-wave background, such as rogue waves \cite{b1,b2} and breathers \cite{b2,b3} in the NLS-MB system have been demonstrated.

However, in contrast to the scalar NLS system, the coupled NLS-MB system possesses some additional system parameters and allows for interaction between different components, which potentially yields rich and significant localized-wave dynamics. Indeed, recent studies indicate that rich nonlinear localized structures do exist in the coupled NLS systems \cite{n0,n1,n2,n3,n4,n5,n6,n7,n8}, such as rogue waves with different structures \cite{n0,n1,n2}, coexistence of different types of localized nonlinear waves \cite{n3,n4,n5,n6}, and so on \cite{n7,n8}.
On the other hand, due to the breaking of the Galilean transformation for the plane-wave background fields, the background frequency plays a key role in the types of nonlinear waves and cannot be ignored in the coupled NLS-MB system. In fact, it is demonstrated recently that the different values of the background frequency in the higher-order NLS model (where the Galilean transformation is broken) can induce different types of localized waves \cite{q1,q2}.
Motivated by these results, we shall study prolific types of nonlinear waves in the NLS-MB system.

In this letter, we present intriguing different types of nonlinear localized and periodic waves in an erbium-doped fiber system, including multi-peak soliton, periodic wave, antidark soliton, and W-shaped soliton (as well as the known bright soliton, rogue wave, and breather). It is found that these waves can be extracted from a unified exact solution under specific parameter conditions. In particular, the multi-peak soliton could be regarded as a single pulse formed by a nonlinear superposition of a soliton and a periodic wave, where each has the same velocity.

\section{NLS-MB system and different types of nonlinear waves}
We consider a resonant erbium-doped fiber system governed by a coupled system of the NLS-MB equations \cite{1,s1,s2}
\begin{eqnarray}
\label{equ:1}
&&E_z=i(\frac{1}{2}E_{tt}+|E|^2 E)+2P ,\nonumber\\
&&P_t=2i\omega P+2E \eta,\nonumber\\
&&\eta_t=-(EP^*+PE^*),
\end{eqnarray}
where $E(z,t)$ is the slowly varying envelope field; $P(z,t)$ is the measure of the polarization of
the resonant medium, which is defined by $P=v_1 v_2^*$; $\eta(z, t)$ denotes the extent of the population inversion, which is given
by $\eta=|v_1|^2-|v_2|^2$, $v_1$ and $v_2$ are the wave functions of the two energy levels of the resonant atoms; $\omega$ is the carrier frequency, and the index $*$ denotes complex conjugate.
In order to study abundant types of nonlinear structures in the NLS-MB model in contrast to the previous results \cite{s1,s2,b1,b2,b3}, we first introduce the following plane-wave background solution with a generalized form
\begin{eqnarray}
&&E_1=ae^{i\theta},~P_1=i k E_1,~\eta_1=\omega k-q k/2,
\end{eqnarray}
where $\theta=q t+\nu z$, $\nu=a^2+2k-q^2/2$, $a$ and $q$ represent the amplitude and frequency of background electric field, respectively, and $k$ is a real parameter which is related to the background amplitude of $P$ component. If the background amplitudes vanish, Eq.(2) reduces to the trivial solution, which can be used to generate standard bright soliton solutions \cite{s1,s2,m1,m2}.
Here we will pay our attention to different types of nonlinear structures in electric field, i.e., the pulse propagation in the $E$ component.
We omit the results in the $P$, $\eta$ components, since their types of nonlinear waves are similar to the ones in the $E$ component with the same initial physical parameters. We present the first-order exact nonlinear wave on the plane-wave solution (2) to reveal rich different types of nonlinear waves.
The construction method is based on the Darboux transformation technique \cite{m3} applied to the Lax pair associated with the NLS-MB model.
For the details, one can construct the solution by solving the partial differential equations (Lax pair) starting from the initial seed solution (2). After that the general first-order exact nonlinear wave solution on the plane-wave background in the $E$ component is given
\begin{equation}
E=E_1\left\{1-\frac{8bm_1[\sin{(\gamma+\mu_1)}-i\sinh{(\beta+i\mu_1)}]}
 {m_3\sin{(\gamma+\mu_2)}-im_2\sinh{(\beta-i\mu_3)}}\right\},
\end{equation}
where
\begin{subequations}
\begin{align}
&\beta=\zeta(t+V_1z),~\gamma=\sigma(t+V_2z),\nonumber\\
&V_1=\upsilon_1+b\sigma\upsilon_2/\zeta,~V_2=\upsilon_1-b\zeta\upsilon_2/\sigma,\nonumber\\
&\upsilon_1=k\omega/(b^2+\omega^2)-q/2,~\upsilon_2=1-k/(b^2+\omega^2),\nonumber\\
&m_1=\sqrt{(i\zeta-\sigma)^2+(2b+iq)^2},\nonumber\\
&m_2=\sqrt{(\alpha_1+\alpha_2)^2-4(2b\zeta+\sigma q)^2},\nonumber\\
&m_3=\sqrt{(\alpha_1-\alpha_2)^2+4(2b\sigma-\zeta q)^2,}\nonumber\\
&\alpha_1=4a^2+4b^2+q^2,~\alpha_2=\zeta^2+\sigma^2,\nonumber\\
&\zeta=\left(\sqrt{\chi^2+16q^2b^2}+\chi\right)^{1/2}/\sqrt{2},\nonumber\\
&\sigma=\left(\sqrt{\chi^2+16q^2b^2}-\chi\right)^{1/2}/\sqrt{2},\nonumber\\
&\chi=4b^2-4a^2-q^2,~\mu_1=\arctan{\left(\frac{2b+iq}{\sigma-i\zeta}\right)},\nonumber\\
&\mu_2=\arctan{\left(\frac{\alpha_1-\alpha_2}{4b\sigma-2q\zeta}\right)},~\mu_3=\arctan{\left(\frac{\alpha_1+\alpha_2}{2iq\sigma-4ib\zeta}\right)}.\nonumber
\end{align}
\end{subequations}

The above expressions depend on the background wave amplitudes $a$, $k$, the background wave frequency $q$, the real parameter $b~(\neq0)$, and the frequency $\omega$. Once the structural parameter $\omega$ is fixed, we are left with four independent parameters $a$, $k$, $q$, $b$.

Remarkably, we find that the unified solution (3) describes abundant different types of nonlinear waves through different choices of the system parameters. For the details, we present a table (Table \uppercase \expandafter {\romannumeral 1}) for the types of nonlinear waves and the corresponding chosen parameter conditions. To our knowledge, some new types of nonlinear waves such as multi-peak soliton, periodic wave, antidark soliton, and W-shaped soliton,
are found in the system for the first time.

To reveal the properties of these new types of nonlinear structures in the system, let us pay our attention to the explicit function expression of the solution (3).
It depends on the hyperbolic functions ($\sinh{\beta}$) and the trigonometric functions ($\sin{\gamma}$), where $\beta$ and $\gamma$ are real functions of $z$ and $t$, and $V_1$, $V_2$ are the corresponding velocities. In this case, the hyperbolic functions and trigonometric functions describe the localization and the periodicity of the transverse distribution $t$ of the nonlinear waves, respectively. Hence this nonlinear structure could be regarded as a single pulse formed by a nonlinear superposition of a soliton and a periodic wave with velocities $V_1$, $V_2$. One should note that the nonlinear wave solution (3) is different from the two-soliton complex solutions \cite{t1,t2} which mix hyperbolic functions of two different (spatial) arguments.
Interestingly, we find that the nonlinear waves described by the unified solution (3) exhibit structural diversity depending on the values of velocity difference, i.e.,  $V_1-V_2$.

\begin{table}[!hbp]
\label{table1}
  \begin{tabular}{|c|c|}
  \hline
  Nonlinear wave type  & Existence condition \\
  \hline
  Breather & $b^2+\omega^2\neq k$, $a\neq b$\\
  \hline
  Rogue wave & $b^2+\omega^2\neq k$, $a=b$ \\
  \hline
  Multi-peak soliton & $b^2+\omega^2=k$, $q\neq0$ \\
  \hline
  Periodic wave &  $b^2+\omega^2=k$, $q=0$, $a^2>b^2$ \\
  \hline
  Antidark soliton & $b^2+\omega^2=k$ , $q=0$, $a^2<b^2$\\
  \hline
  W-shaped soliton & $b^2+\omega^2=k$ , $q=0$, $a=b$\\
  \hline
\end{tabular}
\caption{Types of nonlinear waves in NLS-MB system.}
\end{table}
\begin{figure}[htb]
\centering
\subfigure[]{\includegraphics[height=40mm,width=40mm]{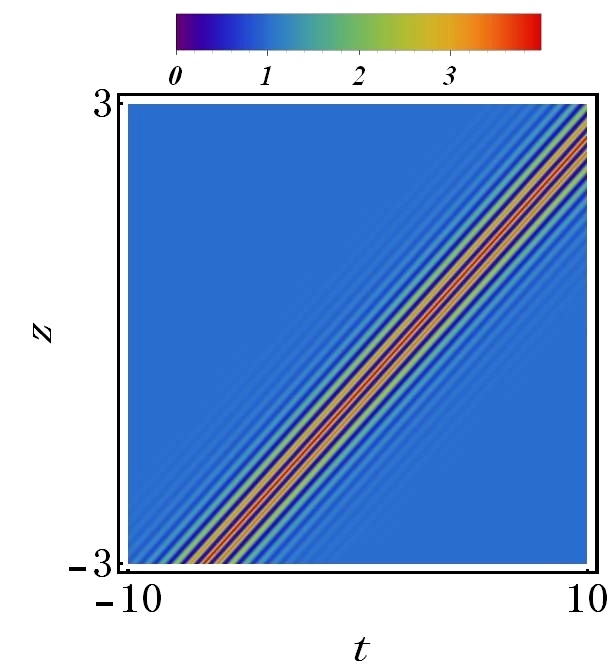}}
\hfil
\subfigure[]{\includegraphics[height=40mm,width=42mm]{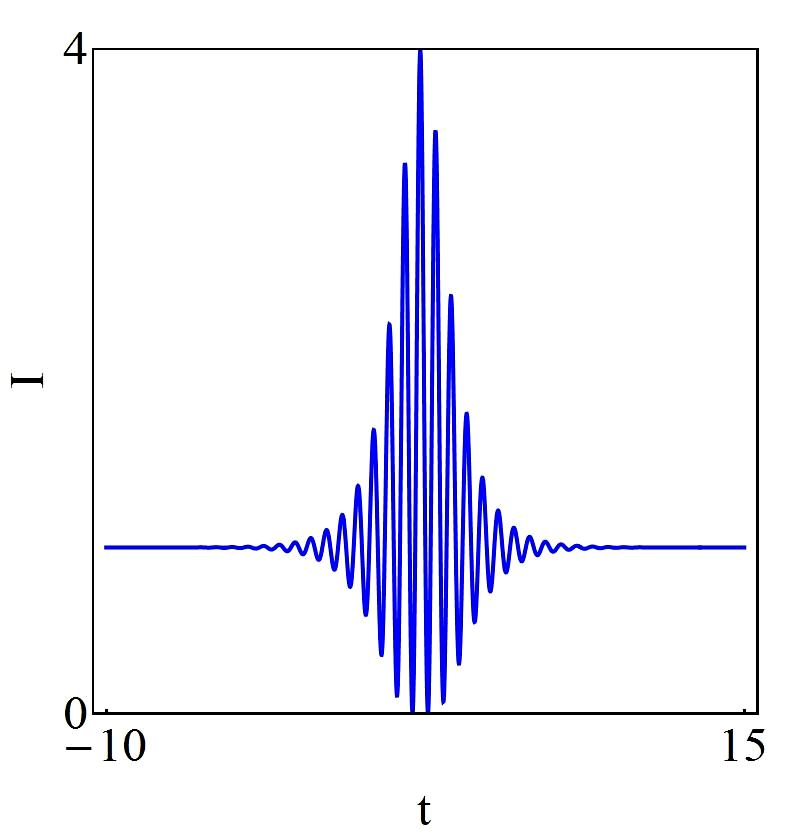}}
\caption{(a) Intensity distributions ($I=|E|^2$) of multi-peak solitons on a plane-wave background, (b) is the profile of (a) at $z=0$. The parameters are $a=1$, $b=0.5$, $\omega=2$, and $q=10$.}
\end{figure}

In the case of nonzero velocity difference, i.e., $V_1\neq V_2$, which implies $\upsilon_2\neq0$ (thus $k\neq b^2+\omega^2$), the expression (3) describes localized waves with breathing behavior on a plane-wave background (i.e., breathers and rogue waves).
Here breathers are the localized breathing waves with a periodic profile in a certain direction; rogue waves are rare, short-lived, and localized in both space and time, which are some special cases of breathers. More specifically, if $q=0$, $\upsilon_2\neq0$, we obtain the Akhmediev breathers \cite{AB} with $a>|b|$, the Kuznetsov-Ma breathers \cite{KMB} with $a<|b|$, and the Peregrine rogue waves \cite{Peregrine} with $a=b$.
We note that the solution (3) includes, as a special case $V_1\neq V_2$, the breather and rogue wave solutions in the NLS-MB system that was reported
in \cite{b1,b2,b3}.

Instead, if $V_1=V_2$, $q\neq0$ (thus $\zeta\neq0$, $\sigma\neq0$), the pulse described by solution (3) is formed by a nonlinear superposition of a soliton and a periodic wave, where each has the same velocity $\upsilon_1$. This result is well depicted in Fig. 1. As expected, in this case, the expression (3) describes a new multi-peak soliton-like pulse propagating along $z$ direction. Namely, the feature of this wave comes from a mixture of a soliton and a periodic wave.
It is interesting, as it looks like higher-order solitons but appears on a plane-wave background.
It should be pointed out that this wave structure is different from the multi-soliton complex \cite{s} formed by a nonlinear superposition of fundamental solitons in coupled NLS systems. We also note that similar multi-peak structures has been observed numerically in the AC-driven damped NLS system \cite{ac1,ac2}. The similarity between the multi-peak NLS-MB solitons reported here, and the AC-driven solitons \cite{ac1,ac2}, is not coincidental. In the limit $\eta \rightarrow 0$, the NLS-MB system (1) reduces to the AC-driven NLS equation.

Additionally, the plane-wave background frequency $q$ has real effects on the structures of multi-peak localized structures, since the corresponding solutions cannot be correlated though the Galileo transformation.  As shown in Fig. 2, we note that, by decreasing the value $|q|$, the numbers of the peaks and the amplitude of the wave decrease.
This implies that one can control the nonlinear excitations via the incident light frequency.
\begin{figure}[htb]
\centering
\includegraphics[height=32mm,width=82mm]{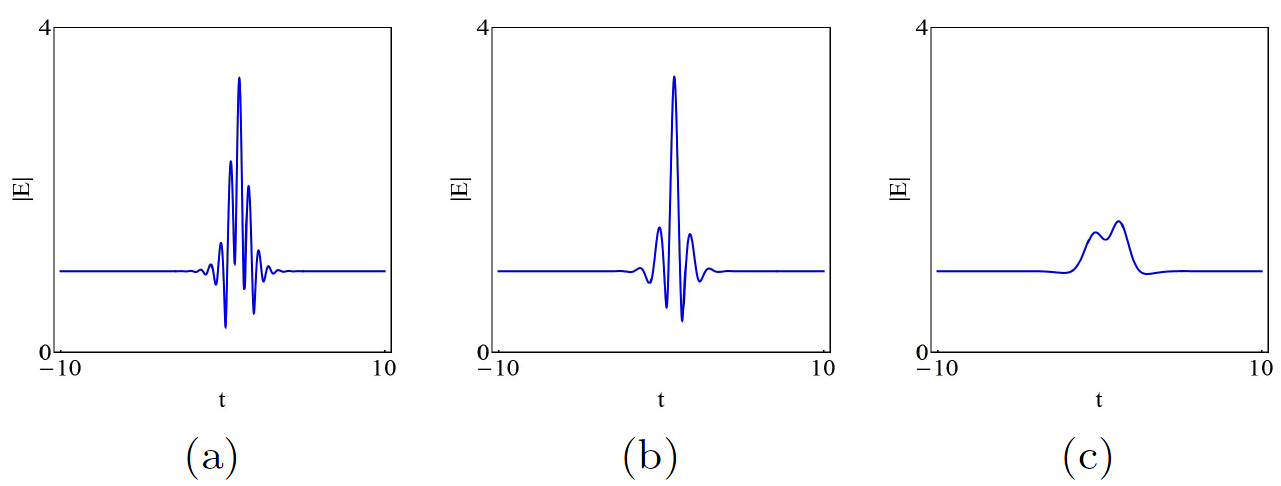}
\caption{Amplitude profiles ($|E|$) of multi-peak solitons on a plane-wave background as the background frequency $q$ decreases, (a) $q=10$, (b) $q=5$, (c) $q=1$. The parameters are $a=1$, $b=1.2$, and $\omega=1$.}
\end{figure}

In order to better understand this multi-peak localized structure of the NLS-MB system (f.i., see the wave in Fig. 1), we will extract separately the periodic wave and soliton from the multi-peak soliton. Specifically, the soliton exist on its own when $\sigma$ vanishes (thus $q=0$, $a^2<b^2$), while the periodic wave exist in isolation when $\zeta$ vanishes (thus $q=0$, $a^2>b^2$). Correspondingly, the explicit expressions read, for the soliton,
\begin{eqnarray}
E_s=E_1\left[\frac{-2s_1^2}{a^2+b^2\cosh{\chi_1}-bs_1\sinh{\chi_1}}-1\right],
\end{eqnarray}
where $\chi_1=2s_1(t+\omega z)$, $s_1=\sqrt{b^2-a^2}$, and for the periodic wave,
\begin{eqnarray}
\label{equ:p}
E_p=E_1\left[\frac{2s_2^2}{a^2+b^2\cos{\chi_2}+bs_2\sin{\chi_2}}-1\right],
\end{eqnarray}
where $\chi_2=2s_2(t+\omega z)$, $s_2=\sqrt{a^2-b^2}$. We then depict the typical intensity distributions
$|E_s|^2$, $|E_p|^2$ in Figs. 3 and 4.

Figure 3 depicts the solitary wave propagating along $z$ direction. It is shown that this soliton lies on a plane-wave background with a peak $|E_s|_{max}^2=(2b-a)^2$, which is called antidark soliton. As is well-known, the existence of antidark solitons is first demonstrated in the scalar NLS system with the third-order dispersion \cite{A}. It is interesting that our result shows that antidark solitons can be induced by the multi-component coupling in the NLS-MB system. We also find that, as $a\rightarrow 0$, this wave will become a standard bright soliton.
\begin{figure}[htb]
\centering
\subfigure[]{\includegraphics[height=34mm,width=40mm]{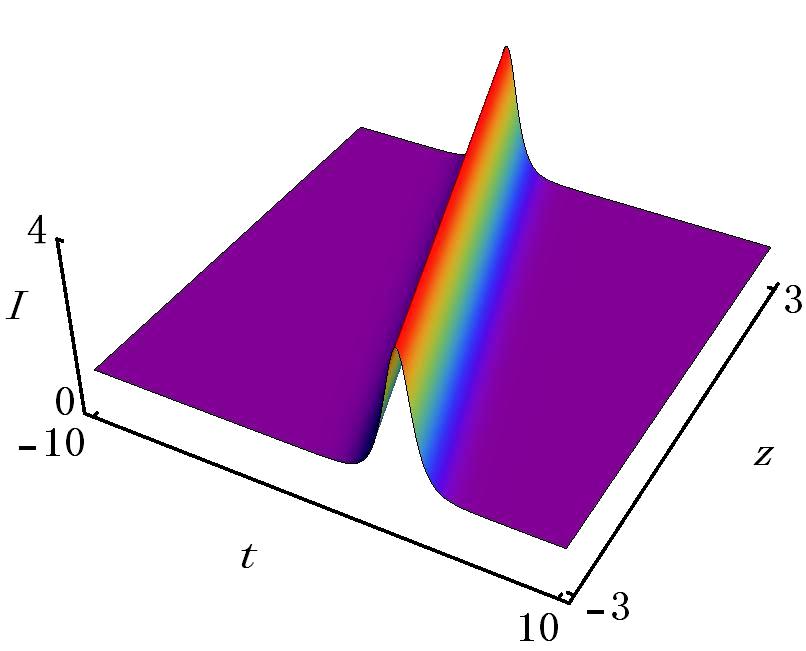}}
\hfil
\subfigure[]{\includegraphics[height=40mm,width=42mm]{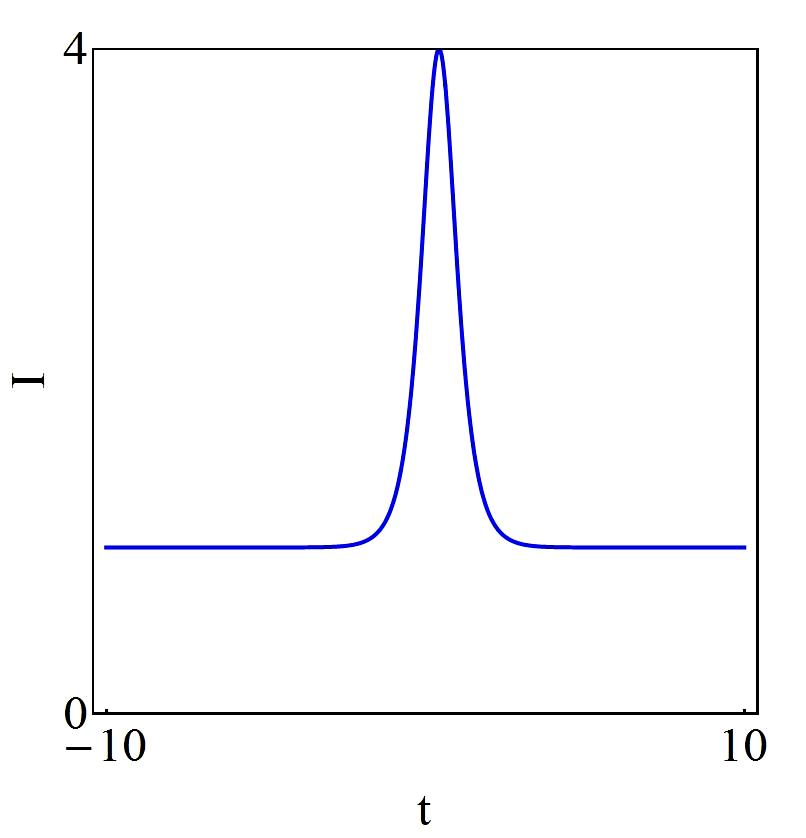}}
\caption{(a) Intensity distributions ($I=|E|^2$) of anti-dark solitons, (b) the profile of (a) at $z=0$. The parameters are $a=1$, $\omega=1$, and $b=1.5$.}
\end{figure}
\begin{figure}[htb]
\centering
\subfigure[]{\includegraphics[height=34mm,width=40mm]{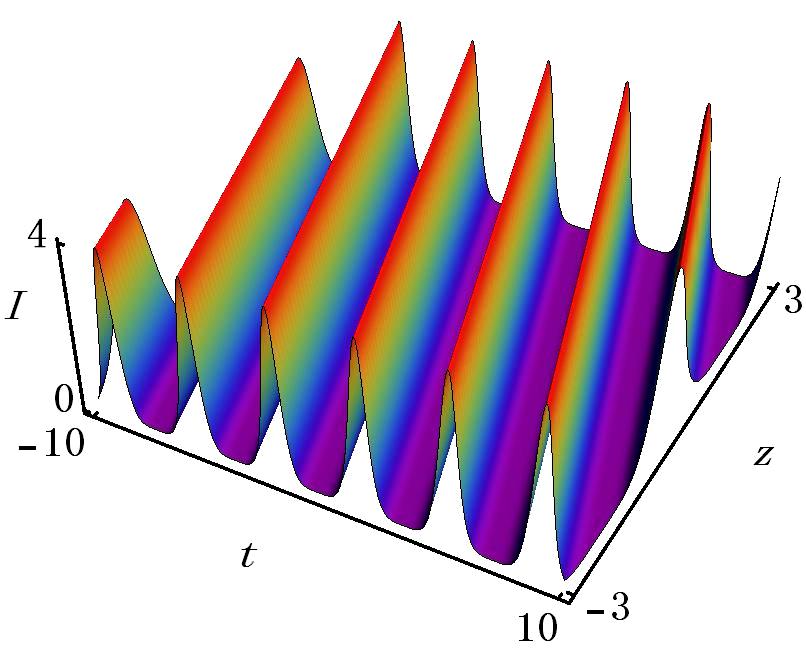}}
\hfil
\subfigure[]{\includegraphics[height=40mm,width=42mm]{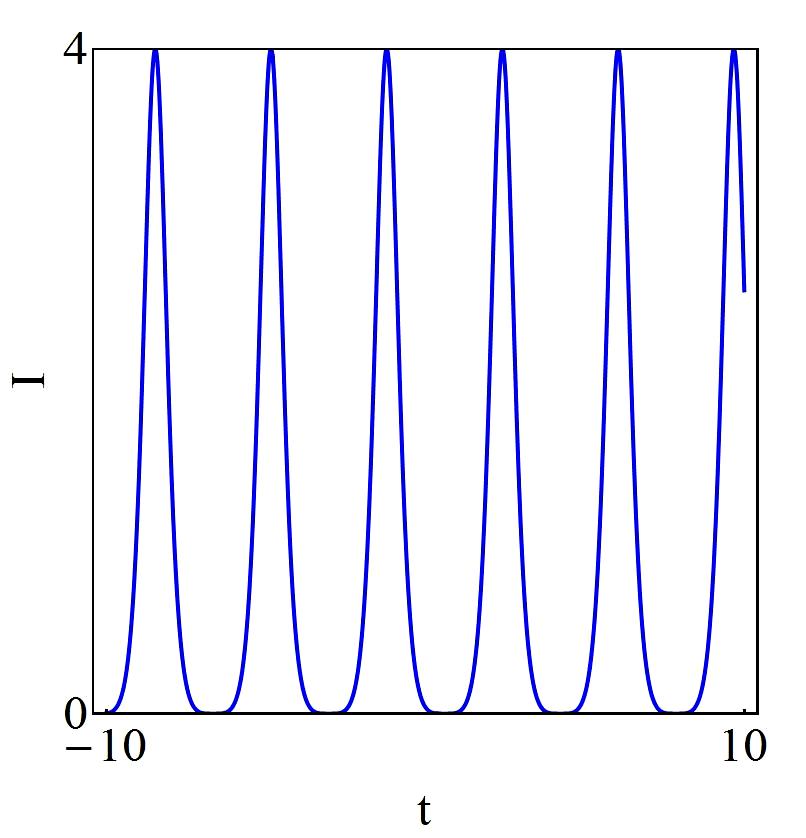}}
\hfil
\subfigure[]{\includegraphics[height=34mm,width=40mm]{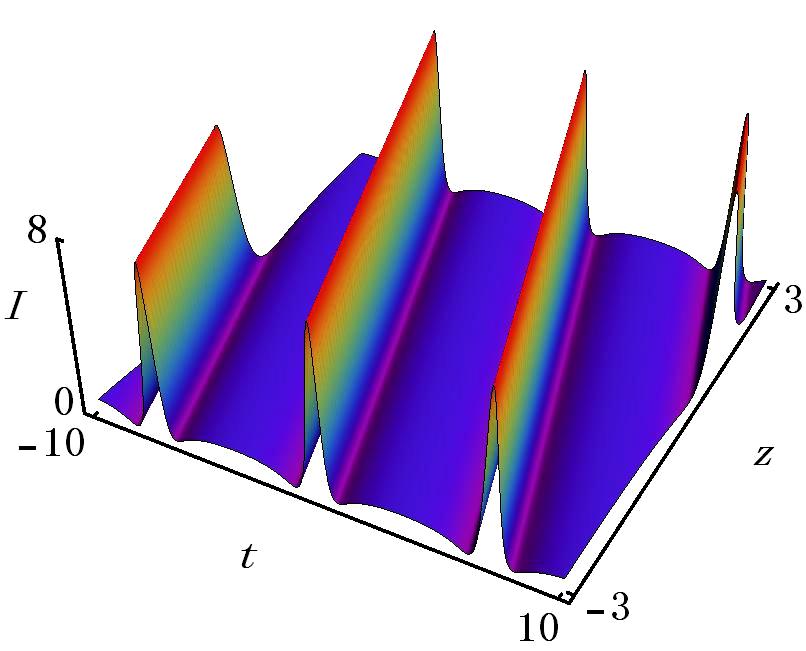}}
\hfil
\subfigure[]{\includegraphics[height=40mm,width=42mm]{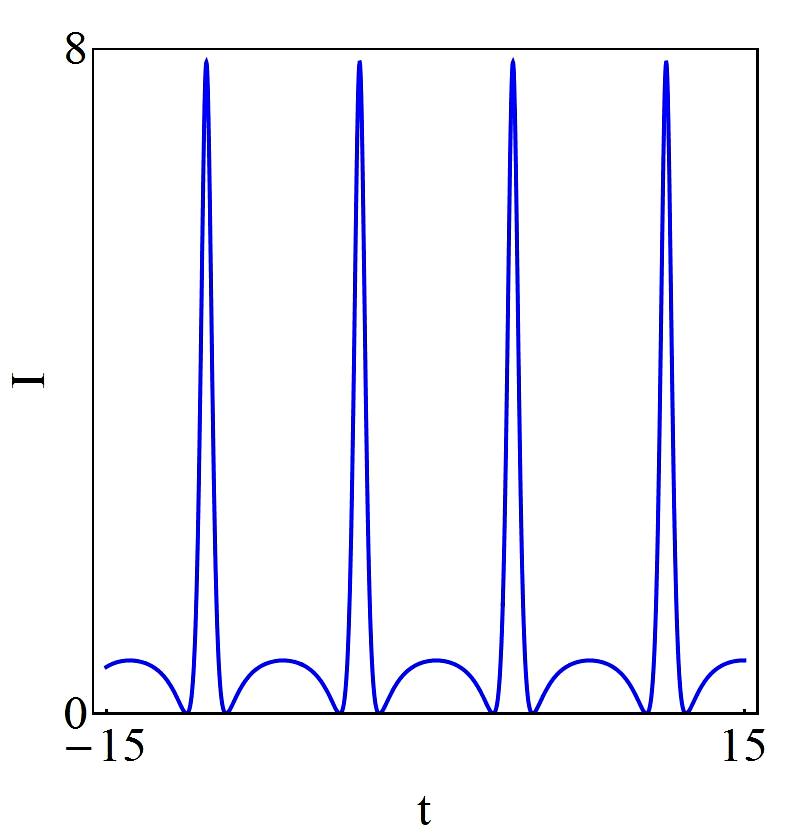}}
\caption{Intensity distributions ($I=|E|^2$) of periodic waves with different structures as the period increases, (a) a periodic wave on a vanishing background with $b=0.5$, (c) a periodic wave on a plane-wave background with $b=0.9$. (b) and (d) are the profiles of (a) and (c) at $z=0$. Other parameters are $a=1$, and $\omega=1$.}
\end{figure}

Figure 4 shows the periodic waves propagating in $z$ direction with the period $D_t=\pi/\sqrt{a^2-b^2}$ along $t$.
We note that, these waves can exhibit different structures as the period $D_t$ increases. Specifically, compared to the
periodic wave on a vanishing background in Fig. 4(a), Figure 4(c) depicts a periodic wave on a plane-wave background, whose
fundamental unit shows a W-shaped soliton structure \cite{w} with two valleys and one peak. It is interesting, as it is a new type of periodic wave
but appears from the same solution. In particular, as the period $D_t\rightarrow\infty$, i.e., $b\rightarrow a$, the periodic wave will
become a single pulse with the W-shaped structure, as shown in Fig. 5. In this case, Eq. (\ref{equ:p}) reduces to the following rational form
\begin{eqnarray}
E_w=E_1\left[1-\frac{4a^2(t+\omega z)^2-4a(t+\omega z)}{2a^2(t+\omega z)^2-2a(t+\omega z)+1}\right].
\end{eqnarray}
The maximum height of the W-shaped wave is nine times the background intensity while the minimum is zero.
Although this property is the same as the one of the Peregrine rogue waves, the W-shaped wave features the soliton-like propagation properties.
It should be pointed out that W-shaped solitons with rational
expressions in scalar NLS systems have been induced by some higher-order effects \cite{q1,q2}. In contrast, the generation of the W-shaped solitons in this system is associated with the interaction between different components.
\begin{figure}[htb]
\centering
\subfigure[]{\includegraphics[height=34mm,width=40mm]{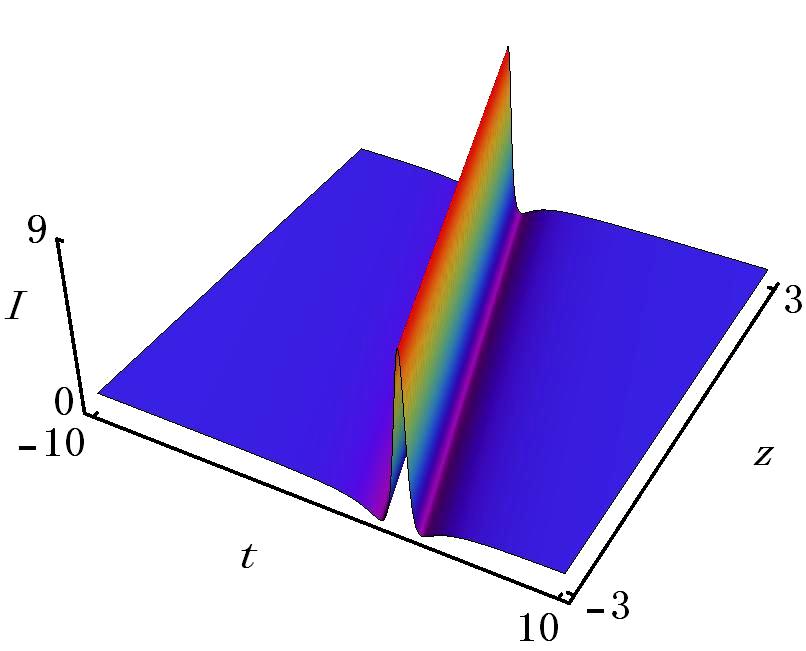}}
\hfil
\subfigure[]{\includegraphics[height=40mm,width=42mm]{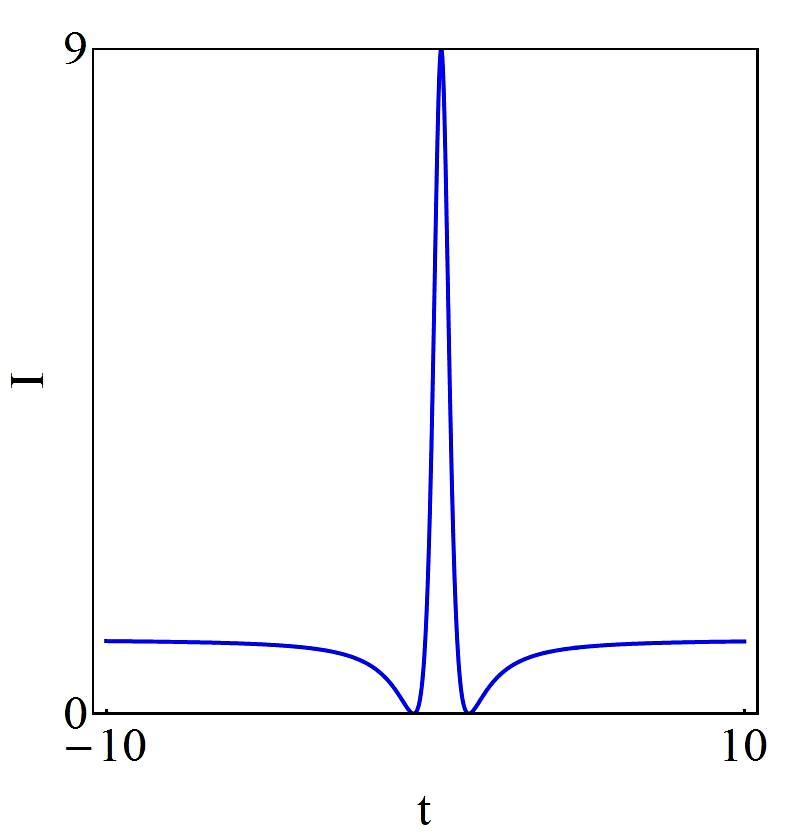}}
\caption{(a) Intensity distributions ($I=|E|^2$) of W-shaped solitons, (b) the profile of (a) at $z=0$. The parameters are $a=1$, $\omega=1$.}
\end{figure}

\section{CONCLUSION}

In summary, we have investigated nonlinear localized and periodic waves on a plane-wave background in an erbium-doped fibre system. Abundant intriguing different types of nonlinear waves, including multi-peak soliton, periodic wave, antidark soliton, and W-shaped soliton (as well as the well-known bright soliton, breather, and rogue wave), have been extracted from the unified exact solution (3). Additionally, our results have shown that the numbers of the peaks of the multi-peak soliton can be controlled by the background frequency. These results demonstrated the structural diversity of nonlinear waves in this system.

On the other hand, our results could provide possibilities to study the nonlinear interactions of different types of nonlinear waves in this system. For example, like recent progresses in \cite{n3,n4,n5,n6,r1,r2,r3}, in which they demonstrated the coexistence and attractive interactions between rogue waves and other types of localized waves,  and the generations of rogue waves from different localized wave backgrounds. Hence it will be interesting to study the nonlinear interactions between rogue waves and other-type waves (such as multi-peak solitons, periodic waves, and anti-dark solitons) by virtue of higher-order solution of the NLS-MB system. Details will be given elsewhere.

\begin{center}
\textbf{ACKNOWLEDGEMENTS}
\end{center}

We are grateful to anonymous reviewers for their valuable comments and Li-Chen Zhao, Ming Zhang, Fa-Kai Wen, Liang Duan for their helpful discussions. This work has been supported by the National Natural Science Foundation of China (NSFC)(Grant No. 11475135), and the ministry of education doctoral program funds (Grant No. 20126101110004).

\end{document}